\documentclass[a4paper,prb,reprint,showkeys,floatfix,superscriptaddress]{revtex4-1}
\pdfoutput=1
\usepackage[latin1]{inputenc}
\usepackage[english]{babel}
\usepackage{amsmath}
\usepackage{ae}
\usepackage{color}
\usepackage{graphicx}
\usepackage{subfigure}
\usepackage{hyperref}
\hypersetup{
  colorlinks   = true, 
  urlcolor     = blue, 
  linkcolor    = blue, 
  citecolor   = blue 
}

\def\vk{\mbox{$\vec{k}$}}

\begin{document}

\title{Current-controlled light scattering and asymmetric plasmon propagation in graphene}

\author{Tobias Wenger}
\affiliation{Department of Microtechnology and Nanoscience (MC2), Chalmers University of Technology, SE-412 96 G\"oteborg, Sweden}
\author{Giovanni Viola}
\affiliation{Department of Microtechnology and Nanoscience (MC2), Chalmers University of Technology, SE-412 96 G\"oteborg, Sweden}
\author{Jari Kinaret}
\affiliation{Department of Physics, Chalmers University of Technology, SE-412 96 G\"oteborg, Sweden}
\author{Mikael Fogelstr\"om}
\affiliation{Department of Microtechnology and Nanoscience (MC2), Chalmers University of Technology, SE-412 96 G\"oteborg, Sweden}
\author{Philippe Tassin}
\affiliation{Department of Physics, Chalmers University of Technology, SE-412 96 G\"oteborg, Sweden}

\date{\today}

\begin{abstract}
We demonstrate that plasmons in graphene can be manipulated using a DC current. A source-drain current lifts the forward/backward degeneracy of the plasmons, creating two modes with different propagation properties parallel and antiparallel to the current. We show that the propagation length of the plasmon propagating parallel to the drift current is enhanced, while the propagation length for the antiparallel plasmon is suppressed. We also investigate the scattering of light off graphene due to the plasmons in a periodic dielectric environment and we find that the plasmon resonance separates in two peaks corresponding to the forward and backward plasmon modes. The narrower linewidth of the forward propagating plasmon may be of interest for refractive index sensing and the DC current control could be used for the modulation of mid-infrared electromagnetic radiation.
\end{abstract}

\maketitle 

\section{Introduction}
Graphene has recently emerged as an exciting tunable plasmonic platform \cite{Koppens2011}, which exhibits large electromagnetic field enhancements \cite{Brar2013} at terahertz to mid-infrared frequencies \cite{Low2014,Grigorenko2012,GarciadeAbajo2014}. The tunability is achieved by applying a gate voltage that allows the charge carrier density in graphene to be controlled externally \cite{Novoselov2004,Ju2011,Gao2013}. This opens up many exciting possibilities for creating tunable photonic devices \cite{Koppens2014,Sun2016,Echtermeyer2016,Shi2015} that can potentially help bridge the gap between photonics and electronics \cite{Schuller2010,Bao2012}. 

Graphene plasmons have also been considered for sensing purposes \cite{Li2014,Rodrigo2015,Marini2015,Farmer2016,Wenger2017} and the large field enhancement can be utilized to sense molecules close to graphene \cite{Li2014}. Such graphene sensors have been shown to be very sensitive to small amounts of molecules and the tunable nature of graphene plasmons enables selective sensing of specific molecules \cite{Rodrigo2015}.

Nonlocal effects of plasmons in graphene, due to the fact that the plasmon wavelength can become comparable with the Fermi wavelength, have previously been discussed in various contexts. The dispersion of long-wavelength plasmons can be treated in a local approximation \cite{Hwang2007,Wunsch2006}. For short-wavelength plasmons, however, it may be important to also consider nonlocal effects to correctly describe the plasmon behaviour \cite{Wang2013,Wenger2016,Jablan2009,Christensen2014}. Recently, experimental verification of the need of a nonlocal approach was achieved in Ref.~\cite{Lundeberg2017}, where graphene plasmons were confined by a metallized nanotip in the direction perpendicular to the graphene sheet. In this article, we use a nonlocal response function to capture these effects. In fact, for the asymmetric plasmon effects that we find, a nonlocal description is necessary. Without nonlocal effects there is nothing that distinguishes between parallel and anti-parallel to the current direction and because of symmetry, plasmons in the two directions behave identically.

\begin{figure}[t]
\centering
\includegraphics[width=0.48\textwidth]{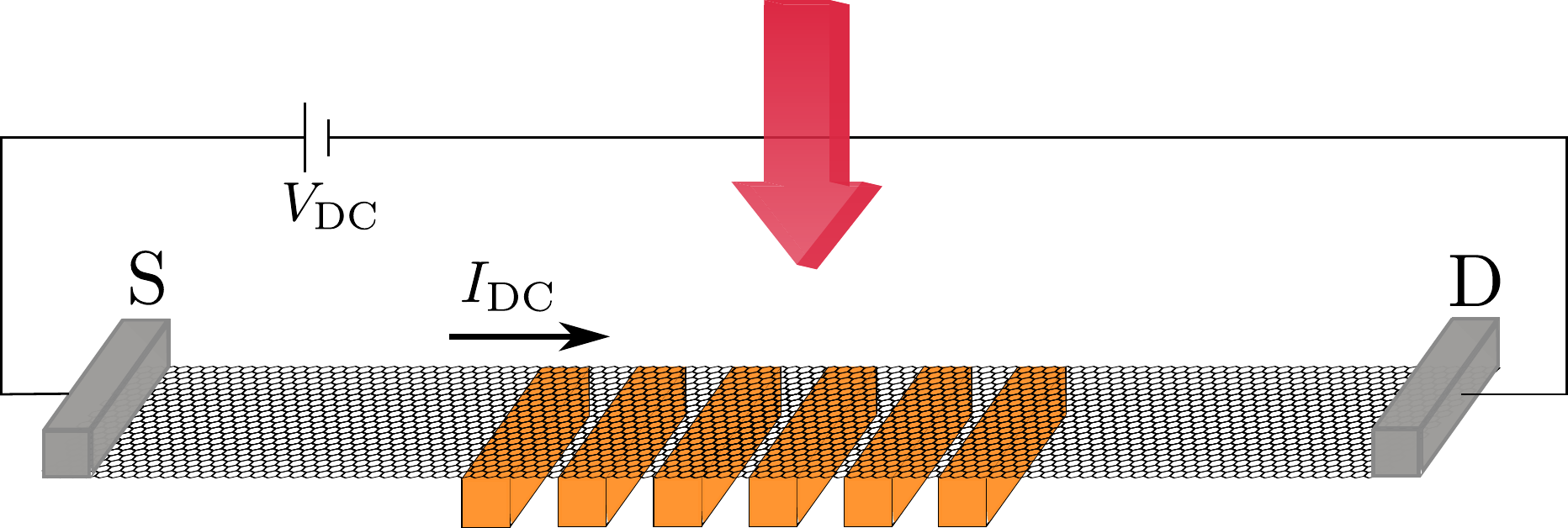}
\caption{\label{fig:setup}A schematic view of the system considered in this article. A source and a drain contact are put far away and a bias voltage can be put over them. This sets up a drift current in the graphene sheet and we study its influence on the properties of graphene plasmons. The plasmons can then be probed by infrared light in the subwavelength grating environment shown in the middle. The grating is in our calculations taken to be infinitely large by use of periodic boundary conditions. The grating we consider has a periodicity of $130$ nm, filling fraction of $0.5$, and the grating material has a dielectric constant $\varepsilon_r=3$. The height of the grating is taken to be $65$ nm.}
\end{figure}

In addition to tuning the carrier concentration in graphene, it is also a possible to induce a stationary current by a voltage bias between a source and a drain contact, see Fig.~\ref{fig:setup}. This may enable even more control over graphene plasmons and by extension also over light at the nanoscale. Indeed, DC currents in graphene have previously been utilized to tune plasmon resonances in metal bow-tie nanoantennas \cite{Emani2012}. 

In this article, we study graphene plasmons in the presence of a constant drift current in graphene. The drift current leads to a lifting of the forward-backward plasmon degeneracy along the current direction and creates asymmetric plasmon propagation. Plasmons propagating in the direction of the electrons are enhanced in terms of propagation length and plasmons propagating against the electrons experience a decreased propagation length. We show that already moderate drift currents allow for control over the graphene plasmon propagation and thus enables electrical control over mid-infrared electromagnetic waves in the subwavelength regime.

Non-equilibrium plasmons in graphene have recently been considered in the literature, such as in Ref.~\cite{Duppen2016}, where the authors studied the plasmon dispersion in the presence of a drift current and found asymmetries in the plasmon dispersion. In Ref.~\cite{Sabbaghi2015} graphene plasmons were investigated for small drift currents and an increased lifetime was found for plasmons propagating parallel to the drift velocity. Furthermore, Ref.~\cite{Sabbaghi2015} found that in the regime of low doping, there can be plasmon gain. Plasmon gain was also reported in Ref.~\cite{Page2015} where graphene was considered to be driven out of equilibrium with a strong laser pulse and the gain was achieved when the electrons relax towards equilibrium, releasing energy as they do so. Graphene out-of equilibrium has also been considered for terahertz amplification \cite{Ghafarian2017}, as well as terahertz emission and detection \cite{Yadav2016,Petrov2017}.

In this article, we investigate graphene plasmons for any drift current. In contrast to previous studies, we also calculate the plasmon propagation lengths, and compute the light scattering by the out-of equilibrium plasmons. We find that the drift-current-induced plasmon control and propagation asymmetries produce clear light-scattering signatures which could be obtained in experiments on gratings \cite{Gao2013,Zhu2013}, ribbons \cite{Ju2011}, or using nanotips \cite{Chen2012,Fei2012}.
distribution.

\begin{figure}[t]
\centering
\includegraphics[width=0.48\textwidth]{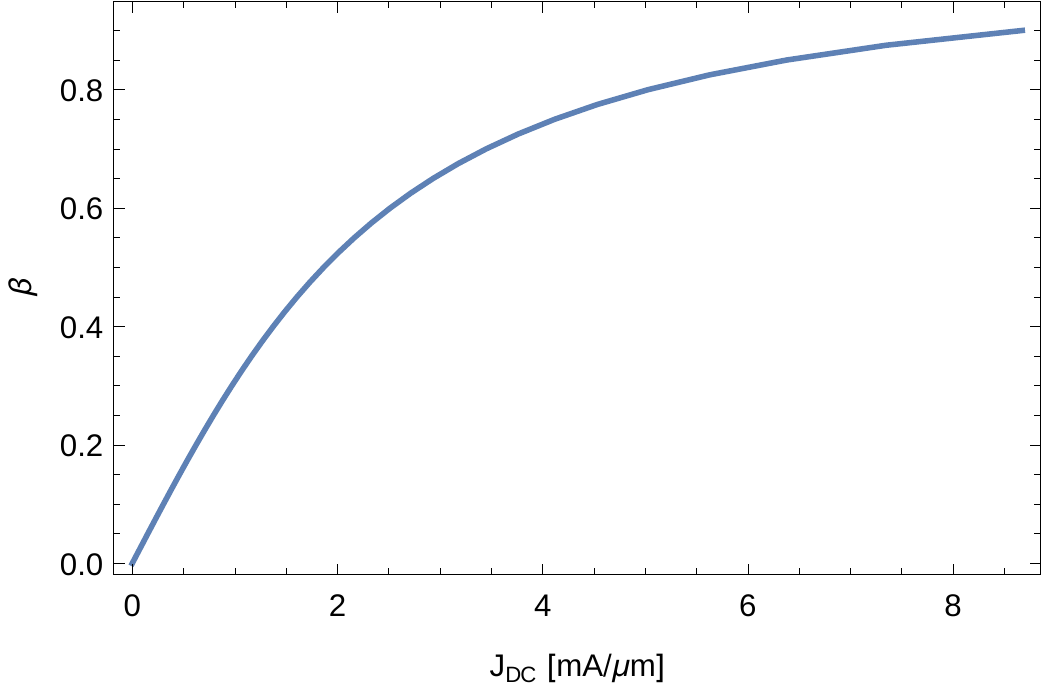}
\caption{\label{fig:current}
The drift velocity $\beta$ as a function of the drift current. The parameter $\beta$ is scaled with the Fermi velocity in graphene. For large values of $\beta$, the current becomes very large because the electron density also increases. Close to $\beta=1$ we do not expect the simple model for the stationary electron distribution to be accurate and we constrain ourselves to $\beta \leq 0.8$ for this reason. The slope of the curve depends on the carrier density and in this article we use $n=1.886\times10^{12}$ cm$^{-2}$ ($E_F=0.16$ eV).}
\end{figure}

\section{Non-equilibrium plasmons}

Graphene transverse magnetic (TM) plasmons in the non-retarded limit satisfy the dispersion equation \cite{Jablan2009,Wenger2016}
\begin{equation}
1+\frac{iq}{\varepsilon_0(\varepsilon_{1}+\varepsilon_{2})\omega}\sigma(q,\omega)=0
\end{equation}
where $\sigma(q,\omega)$ is the nonlocal sheet conductivity of graphene, and $q=q_1+iq_2$ is the complex wave vector. Expanding this equation for small $q_2/q_1$, i.e., small plasmon losses, yields 

\begin{align}
\varepsilon_1+\varepsilon_2+\frac{q_1\sigma_2(q_1,\omega)}{\omega\varepsilon_0} =0\label{eq:dispReal}\\
\frac{q_2}{q_1}=\frac{\sigma_1(q_1,\omega)}{\frac{\partial}{\partial q_1}(q_1\sigma_2(q_1,\omega))},\label{eq:dispImag}
\end{align}
where $\sigma(q_1,\omega)=\sigma_1(q_1,\omega)+i\sigma_2(q_1,\omega)$. The first of these equations can be solved for a given $\omega$ to obtain $q_1$, which can be inserted in the second equation to give $q_2$, which is a measure of the plasmon damping. The conductivity of graphene can be obtained from a linear response calculation of the polarizability \cite{Hwang2007,Wunsch2006}
\begin{align}
\Pi(q,\omega)&=\lim_{\eta\to 0^+}\frac{g_sg_v}{2}\int \frac{d^2\vec{k}}{(2\pi)^2} \\
&\times\sum_{\lambda,\lambda'=+,-}\frac{f_{\lambda,k}-f_{\lambda',k'}}{\omega +i\eta -\lambda'v_F |\vk+\vec{q}| + \lambda v_F k}\\
&\times(1+\lambda\lambda'\cos\phi_{k,k'}).\label{eq:Pi}
\end{align}
where $\vec{k}'=\vk+\vec{q}$, and $\phi_{k,k'}=\phi_{k+q}-\phi_k$. The conductivity can then be treated using Mermin's approach to include a phenomenological relaxation time $\tau$ in a number-conserving way \cite{Mermin1970,Jablan2009}:
\begin{equation}\label{eq:cond}
\sigma(q,\omega)=\frac{ie^2\omega}{q^2}\frac{(1+\frac{i\Gamma}{\omega}) \Pi(q,\omega+i\Gamma)}{1+\frac{i\Gamma}{\omega}\frac{\Pi(q,\omega+i\Gamma)}{\Pi(q,0)}}.
\end{equation}
The phenomenological parameter $\Gamma=\tau^{-1}$ accounts for intraband scattering that is known to exist from experiments and to cause significantly increased plasmon damping \cite{Jablan2009}. We take $\tau=170$ fs, which is achieved in experiments \cite{Efetov2010,Tassin2013}. The plasmon propagation distance in units of the plasmon wavelength can be obtained by $L_p/\lambda_p=q_1/(4\pi q_2)$ \cite{peres,Viola2017}, which gives the number of oscillations the plasmon makes before it decays.\\

The stationary electron distribution in the presence of relaxation and a drift velocity can be written as \cite{gantmakher}
\begin{equation}\label{eq:distribution}
f_{\lambda,k}=\frac{1}{1+e^{(\lambda E_k-v_F\vec{\beta}\cdot \vec{k}-\mu)/(k_B T)}},
\end{equation}
where $\vec{\beta}$ is the drift velocity vector scaled with $v_F$ and $E_k=v_F|k|$ is the linear dispersion for Dirac electrons. When $|\vec{\beta}|=0$, Eq.~\eqref{eq:distribution} reduces to the ordinary Fermi distribution. Throughout the article we take $\vec{\beta}$ to be either parallel or anti-parallel to the plasmon wave vector $\vec{q}$. Eq.~\eqref{eq:distribution} describes an asymmetric electron distribution where the asymmetry is parametrized by the parameter $|\vec{\beta}|$, which takes values between $+1$ and $-1$. The distribution given by Eq.~\eqref{eq:distribution} does not conserve the number of particles as $\beta$ increases from zero. Since the system we consider is an open system, we take the approach that the leads set the chemical potential and particle number. A similar approach was also taken in Ref.~\cite{Duppen2016}. At zero temperature, the DC current can be calculated using the formula
\begin{equation}
J_{\text{DC}}=-eg_sg_v\int \frac{d^2\vec{k}}{(2\pi)^2} \frac{\vec{k}}{|\vec{k}|} f_{+,k}, 
\end{equation}
where $f_{+,k}$ is the conduction band distribution from Eq.~\eqref{eq:distribution}. The relationship between the obtained current and the parameter $\beta$ is shown in Fig.~\ref{fig:current}, for $E_F=0.16$ eV. We continue to work at zero temperature throughout the rest of the article.

\begin{figure}[t]
\centering
\includegraphics[width=0.48\textwidth]{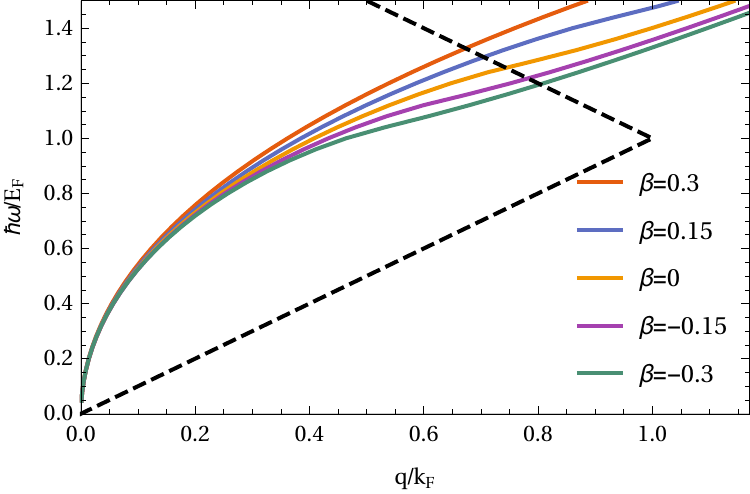}
\caption{\label{fig:disp} Plasmon dispersion for various drift currents. For $\beta > 0$ ($\beta<0$), the electrons propagate parallel (anti-parallel) to the plasmons. Plasmons propagating parallel to the drift velocity are blue-shifted and obtain a higher group velocity. Vice-versa, the plasmons propagating anti-parallel to the drift velocity are red shifted and acquire a lower group velocity. These effects are increasing with increasing magnitude of the drift velocity.}
\end{figure}

Fig.~\ref{fig:disp} shows the plasmon dispersion relation obtained by solving Eq.~\eqref{eq:dispReal} for various drift velocities. The figure shows only plasmons propagating in the positive $x$ direction and the plasmons are affected differently depending on if the drift velocity is in the positive or negative $x$ direction. The plasmons may propagate in either positive or negative direction and the drift current lifts the degeneracy of the forward and backward plasmon modes. The difference between the propagation directions grows with increasing drift velocity and the effect is larger for the high-energy plasmons as can be seen in Fig.~\ref{fig:disp}. This asymmetric behavior was previously noticed in Ref.~\cite{Duppen2016}.

\begin{figure}[t]
\centering
\includegraphics[width=0.48\textwidth]{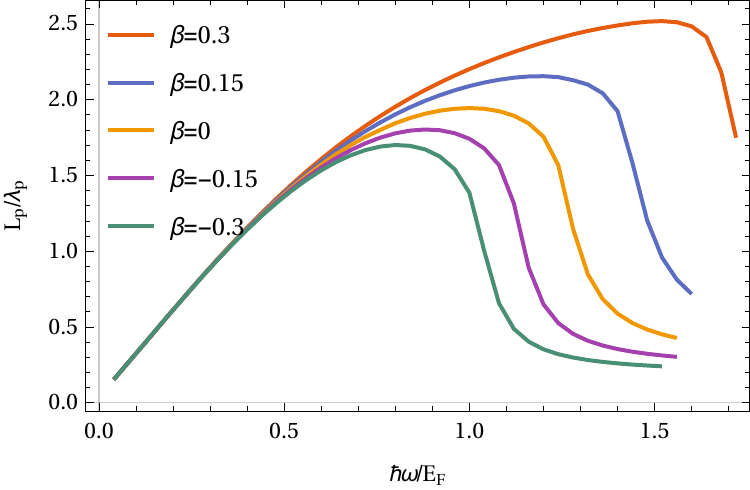}
\caption{\label{fig:prop} The plasmon propagation distance for various drift velocities. At energies around the Fermi energy the propagation is enhanced for plasmons propagating parallel to the drift velocity and it is reduced for plasmons propagating anti-parallel to the drift velocity. For plasmon energies larger than the Fermi energy, the difference between parallel and anti-parallel propagation is very large.}
\end{figure}

We now turn our attention to the plasmon propagation and how it is affected by the drift current. Fig.~\ref{fig:prop} shows the plasmon propagation length as a function of the plasmon energy for different values of the drift velocity $\beta$. Plasmons propagating parallel to the drift velocity experience an enhanced propagation length, whereas plasmons propagating anti-parallel to the drift velocity have a suppressed propagation. This effect allows for external control and enhancement of the plasmon propagation without increasing the mobility of the graphene samples. In addition, the region where the effect of the drift current is most pronounced --- the high-energy region above $E_F$ --- is the region with the strongest localization and is typically heavily damped \cite{Yan2013,peres}. These highly localized plasmons have the potential to facilitate strong coupling to emitters in its vicinity \cite{Koppens2011} and the drift current supplies a method by which their damping can be suppressed. However, care must be taken to not heat the graphene sample extensively as this leads to larger plasmon losses and, hence, smaller propagation lengths.

\section{Light scattering from non-equilibrium plasmons}
We next consider the effect of a drift current on the optical properties of graphene. As shown before \cite{Ju2011,Zhu2013,Gao2013}, these optical properties are intimately linked with plasmons, and we evaluate them using COMSOL, a finite element method (FEM) solver. Graphene enters the solver as a conducting boundary condition where the sheet current induced by the incident field is given by
\begin{equation}
j(x,\omega)=\int \sigma(x-x',\omega)E(x',\omega)\ dx' ,
\end{equation}
where the integration is over the graphene sheet and the electric field is the in-plane component along the graphene sheet. By considering the grating region to be infinite in extension, the system can be simulated using a unit cell containing one period of the grating and periodic boundary conditions. The space-dependent graphene conductivity is calculated from the nonlocal conductivity in Eq.~\eqref{eq:cond} by
\begin{equation}
\sigma(x-x',\omega)=\sum_n e^{ik_n(x-x')}\sigma(k_n,\omega),
\end{equation} 
where $k_n=2\pi n/d$ and $d$ is the periodicity introduced in the system by the subwavelength grating. The grating is needed to overcome the large momentum mismatch between the plasmons and the incident light. Without the grating, the response of graphene to the incident field would not show any signs of plasmons. For the parameters we have chosen, $k_1$ corresponds to $k_1/k_F=0.2$ in Fig.~\ref{fig:disp}.

\begin{figure}[t]
\centering
\includegraphics[width=0.48\textwidth]{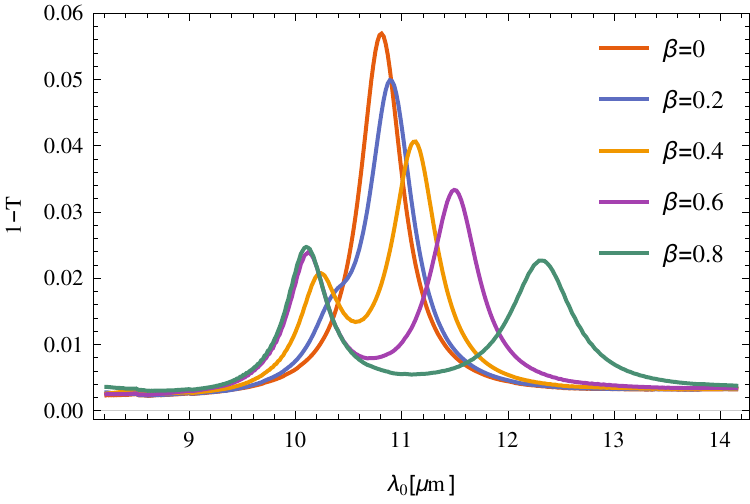}
\caption{\label{fig:extinct}Extinction curves as a function of incident wavelength for various values of $\beta$. As $\beta$ is increased the plasmon peak separates into two distinguishable modes, related with plasmons propagating parallel and anti-parallel to the drifting electrons. The separation of the modes grows as the magnitude of $\beta$ grows and for values around $\beta\geq 0.4$, the two peaks are visible.}
\end{figure}

The system we consider here consists of a graphene sheet with Fermi energy $E_F=0.16$ eV and a dielectric grating with periodicity $d=130$ nm, see Fig.~\ref{fig:setup}. Using the method described above, we compute the light extinction i.e., unity minus transmission, through the system for various values of the drift velocity $\beta$. The results are shown in Fig.~\ref{fig:extinct} and clear plasmon peaks appear when the incident light is resonant with the plasmon with wavelength equal to the grating periodicity. As expected from Fig.~\ref{fig:disp}, the lifting of the forward-backward degeneracy makes the plasmon peak separate into two distinguishable peaks as the drift current is increased. Fig.~\ref{fig:extinct} shows that the separation of the extinction peaks grows with increasing $\beta$ and the high energy mode (smaller incident wavelength) is narrower in agreement with the analysis of the propagation length. The narrower peak is a signature of plasmons propagating parallel to the drifting electrons and the wider peak is a signature of the plasmons propagating anti-parallel to the drifting electrons.

\section{Discussion}
The effects of the drift velocity on the plasmon propagation allows not only for increased propagation distances in experimental settings, but also creates new possibilities for creating plasmonic graphene devices with active control over the plasmons. The current creates a region of essentially unidirectional plasmon propagation that can be turned on and off by the current and the direction can also be reversed by reversing the current. This can potentially be used to create plasmonic modulators where the DC electric current modulates the plasmon propagation and thus the plasmon signal. This could be lead to compact photonic devices, with sizes comparable to the plasmon wavelength on the order of $100$ nm, in the mid-infrared.

Direct measurements of the enhanced plasmon propagation distance should be possible using nanotip experiments similar to Refs.~\cite{Chen2012,Fei2012}. In these experiments, only one nanotip was used and the plasmons were launched from the tip, reflected off a boundary, and returned to the tip for out-coupling to a photon detector. In such a setup, the plasmon necessarily travels both parallel and anti-parallel to the induced current making the results more difficult to interpret. However, since the effects of propagation enhancement and increased losses are not each other's inverse, it should still be possible to observe the effects of the DC current in a one-tip experiment. We point out, however, that a two-tip experiment, where the plasmon is launched at one tip and outcoupled at the other, would provide a measurement signal that is more directly related to the results obtained in this article.

The mechanism discussed here could be important for the modulation of mid-infrared signals, since the drift control allows control over the extinction. It can also be used in experiments where control over the linewidth is desired.

\section{Conclusions}
We have shown that graphene plasmons can be controlled by a DC current in the graphene sheet. This leads to asymmetric plasmon propagation and, in particular, plasmons propagating in the same direction as the electrons in the current experience an enhanced propagation length. We find that optical signatures of this asymmetry can be detected by light scattering in a subwavelength grating environment, where the equilibrium plasmon peak separates into two separate peaks as the DC current increases. The two peaks correspond to plasmons propagating parallel and anti-parallel to the current.

Furthermore, we discussed the implications of asymmetric plasmon propagation. This leads to narrower linewidths of the plasmons, which could be useful for sensing applications utilizing graphene plasmons. The ability to control terahertz and mid-infrared wavelengths using DC current control over graphene plasmons could enable compact modulators at these wavelengths. 

The effects we discussed in this article can be measured directly in optical scattering experiments, or, alternatively, the asymmetric plasmon propagation can be measured using nanotips.

\bibliography{refs}

\end{document}